\pdfoutput=1
\documentclass[12pt,a4paper]{article}
\usepackage{ifthen} % for conditional statements
\usepackage{subfigure}
\usepackage{graphics}
\usepackage{makecell}
\usepackage[figuresright]{rotating}
\newboolean{pdflatex}
\setboolean{pdflatex}{true} % False for eps figures 
\newboolean{articletitles}
\setboolean{articletitles}{true} % False removes titles in references
\newboolean{uprightparticles}
\setboolean{uprightparticles}{false} %True for upright particle symbols
\def\paperauthors{LHCb collaboration} % Leave as is for PAPER, CONF and FIGURE
\def\paperasciititle{Observation of new Omegac states decaying to the XicK final state} % Set ASCII title here !! MAKE sure it's only ASCII characters !! 
\def\papertitle{Observation of new $\POmega_c^{0}$ states decaying to the $\PXi_c^+K^-$ final state} % Latex formatted title
\def\paperkeywords{{LHCb},{Flavour physics}} % Comma separated list
\def\papercopyright{\the\year\ CERN for the benefit of the LHCb collaboration} % new since 9/Apr/2018
\def\paperlicence{CC BY 4.0 licence}
\def\paperlicenceurl{https://creativecommons.org/licenses/by/4.0/}
\usepackage[top=1in, bottom=1.25in, left=1in, right=1in]{geometry}

\usepackage{microtype}
\usepackage{lineno}  % for line numbering during review
\usepackage{xspace} % To avoid problems with missing or double spaces after
\usepackage{caption} %these three command get the figure and table captions automatically small

\usepackage{graphicx}  % to include figures (can also use other packages)
\usepackage{color}
\usepackage{colortbl}
\graphicspath{{./}} % Make Latex search fig subdir for figures
\usepackage{amsmath} % Adds a large collection of math symbols
\usepackage{amssymb}
\usepackage{amsfonts}
\usepackage{upgreek} % Adds in support for greek letters in roman typeset

\newcommand*\patchAmsMathEnvironmentForLineno[1]{%
\expandafter\let\csname old#1\expandafter\endcsname\csname #1\endcsname
\expandafter\let\csname oldend#1\expandafter\endcsname\csname
end#1\endcsname
 \renewenvironment{#1}%
   {\linenomath\csname old#1\endcsname}%
   {\csname oldend#1\endcsname\endlinenomath}%
}
\newcommand*\patchBothAmsMathEnvironmentsForLineno[1]{%
  \patchAmsMathEnvironmentForLineno{#1}%
  \patchAmsMathEnvironmentForLineno{#1*}%
}
\AtBeginDocument{%
\patchBothAmsMathEnvironmentsForLineno{equation}%
\patchBothAmsMathEnvironmentsForLineno{align}%
\patchBothAmsMathEnvironmentsForLineno{flalign}%
\patchBothAmsMathEnvironmentsForLineno{alignat}%
\patchBothAmsMathEnvironmentsForLineno{gather}%
\patchBothAmsMathEnvironmentsForLineno{multline}%
\patchBothAmsMathEnvironmentsForLineno{eqnarray}%
}

\usepackage{hyperxmp}

\usepackage[pdftex,
            pdfauthor={\paperauthors},
            pdftitle={\paperasciititle},
            pdfkeywords={\paperkeywords},
            pdfcopyright={Copyright (C) \papercopyright},
            pdflicenseurl={\paperlicenceurl}]{hyperref}
\usepackage[colorinlistoftodos,textsize=scriptsize]{todonotes}

\usepackage[bottom,flushmargin,hang,multiple]{footmisc}

\usepackage[all]{hypcap} % Internal hyperlinks to floats.

\usepackage{xspace} 
\usepackage{upgreek}

\def\lhcb   {\mbox{LHCb}\xspace}

\def\MagUp {\mbox{\em Mag\kern -0.05em Up}\xspace}

\ifthenelse{\boolean{uprightparticles}}%
{

 \def\Ppi         {\ensuremath{\uppi}\xspace}

 \def\PDelta      {\ensuremath{\Delta}\xspace}                 
 \def\PXi         {\ensuremath{\Xi}\xspace}                 
 \def\PLambda     {\ensuremath{\Lambda}\xspace}                 
 \def\PSigma      {\ensuremath{\Sigma}\xspace}                 
 \def\POmega      {\ensuremath{\Omega}\xspace}                 
 \def\PUpsilon    {\ensuremath{\Upsilon}\xspace}
 \let\oldPi\Pi
 \def\PPi         {\ensuremath{\oldPi}\xspace}

 \def\PB      {\ensuremath{\mathrm{B}}\xspace}                 
                  
 \def\PD      {\ensuremath{\mathrm{D}}\xspace}

 \def\PK      {\ensuremath{\mathrm{K}}\xspace}

 \def\Pb      {\ensuremath{\mathrm{b}}\xspace}                 
 \def\Pc      {\ensuremath{\mathrm{c}}\xspace}

 \def\Pi      {\ensuremath{\mathrm{i}}\xspace}

 \def\Pp      {\ensuremath{\mathrm{p}}\xspace}                 
 \def\Pq      {\ensuremath{\mathrm{q}}\xspace}                 
                  
 \def\Ps      {\ensuremath{\mathrm{s}}\xspace}

 \def\thebaroffset{0.0em}
}
{

 \def\Ppi         {\ensuremath{\pi}\xspace}

 \mathchardef\PDelta="7101
 \mathchardef\PXi="7104
 \mathchardef\PLambda="7103
 \mathchardef\PSigma="7106
 \mathchardef\POmega="710A
 \mathchardef\PUpsilon="7107
 \mathchardef\PPi="7105
                  
 \def\PB      {\ensuremath{B}\xspace}                 
                  
 \def\PD      {\ensuremath{D}\xspace}

 \def\PK      {\ensuremath{K}\xspace}

 \def\Pb      {\ensuremath{b}\xspace}                 
 \def\Pc      {\ensuremath{c}\xspace}

 \def\Pi      {\ensuremath{i}\xspace}

 \def\Pp      {\ensuremath{p}\xspace}                 
 \def\Pq      {\ensuremath{q}\xspace}                 
                  
 \def\Ps      {\ensuremath{s}\xspace}

 \def\thebaroffset{0.18em}
}
\newcommand{\offsetoverline}[2][\thebaroffset]{\kern #1\overline{\kern -#1 #2}}%

\makeatletter
\ifcase \@ptsize \relax% 10pt
  \newcommand{\miniscule}{\@setfontsize\miniscule{4}{5}}% \tiny: 5/6
\or% 11pt
  \newcommand{\miniscule}{\@setfontsize\miniscule{5}{6}}% \tiny: 6/7
\or% 12pt
  \newcommand{\miniscule}{\@setfontsize\miniscule{5}{6}}% \tiny: 6/7
\fi
\makeatother

\DeclareRobustCommand{\optbar}[1]{\shortstack{{\miniscule (\rule[.5ex]{1.25em}{.18mm})}
  \\ [-.7ex] $#1$}}

\def\quark     {{\ensuremath{\Pq}}\xspace}
\def\quarkbar  {{\ensuremath{\overline \quark}}\xspace}
\def\qqbar     {{\ensuremath{\quark\quarkbar}}\xspace}

\def\squark    {{\ensuremath{\Ps}}\xspace}

\def\cquark    {{\ensuremath{\Pc}}\xspace}

\def\bquark    {{\ensuremath{\Pb}}\xspace}

\def\pion   {{\ensuremath{\Ppi}}\xspace}
\def\piz    {{\ensuremath{\pion^0}}\xspace}
\def\pip    {{\ensuremath{\pion^+}}\xspace}

\def\kaon    {{\ensuremath{\PK}}\xspace}
\def\KorKbar {\kern \thebaroffset\optbar{\kern -\thebaroffset \PK}{}\xspace}

\def\Km      {{\ensuremath{\kaon^-}}\xspace}

\def\D       {{\ensuremath{\PD}}\xspace}

\def\DorDbar {\kern \thebaroffset\optbar{\kern -\thebaroffset \PD}\xspace}

\def\Dp      {{\ensuremath{\D^+}}\xspace}
\def\Dm      {{\ensuremath{\D^-}}\xspace}

\def\DpDm    {\ensuremath{\Dp {\kern -0.16em \Dm}}\xspace}

\def\B       {{\ensuremath{\PB}}\xspace}

\def\BorBbar {\kern \thebaroffset\optbar{\kern -\thebaroffset \PB}\xspace}

\def\Bd      {{\ensuremath{\B^0}}\xspace}

\def\BdorBdbar {\kern \thebaroffset\optbar{\kern -\thebaroffset \Bd}\xspace}

\def\Bs      {{\ensuremath{\B^0_\squark}}\xspace}

\def\BsorBsbar {\kern \thebaroffset\optbar{\kern -\thebaroffset \Bs}\xspace}

\def\Y#1S{\ensuremath{\PUpsilon{(#1S)}}\xspace}

\def\proton      {{\ensuremath{\Pp}}\xspace}

\def\Lz          {{\ensuremath{\PLambda}}\xspace}

\def\LorLbar     {\kern \thebaroffset\optbar{\kern -\thebaroffset \PLambda}\xspace}

\def\Xires       {{\ensuremath{\PXi}}\xspace}

\def\Omegares    {{\ensuremath{\POmega}}\xspace}

\def\Xic         {{\ensuremath{\Xires_\cquark}}\xspace}

\def\Xicp        {{\ensuremath{\Xires^+_\cquark}}\xspace}

\def\Omegac      {{\ensuremath{\Omegares^0_\cquark}}\xspace}

\def\Lb           {{\ensuremath{\Lz^0_\bquark}}\xspace}

\def\Xibm         {{\ensuremath{\Xires^-_\bquark}}\xspace}

\def\Omegab       {{\ensuremath{\Omegares^-_\bquark}}\xspace}

\def\to                 {\ensuremath{\rightarrow}\xspace}

\def\AT#1     {\ensuremath{A_{\mathrm{T}}^{#1}}\xspace}           % 2

\def\C#1      {\ensuremath{\mathcal{C}_{#1}}\xspace}                       % 9
\def\Cp#1     {\ensuremath{\mathcal{C}_{#1}^{'}}\xspace}                    % 7
\def\Ceff#1   {\ensuremath{\mathcal{C}_{#1}^{\mathrm{(eff)}}}\xspace}        % 9  
\def\Cpeff#1  {\ensuremath{\mathcal{C}_{#1}^{'\mathrm{(eff)}}}\xspace}       % 7
\def\Ope#1    {\ensuremath{\mathcal{O}_{#1}}\xspace}                       % 2
\def\Opep#1   {\ensuremath{\mathcal{O}_{#1}^{'}}\xspace}                    % 7

\newcommand{\aunit}[1]{\ensuremath{\text{\,#1}}}       
\newcommand{\tev}{\aunit{Te\kern -0.1em V}\xspace}
\newcommand{\gev}{\aunit{Ge\kern -0.1em V}\xspace}
\newcommand{\mev}{\aunit{Me\kern -0.1em V}\xspace}
\newcommand{\kev}{\aunit{ke\kern -0.1em V}\xspace}
\newcommand{\ev}{\aunit{e\kern -0.1em V}\xspace}
 
\newcommand{\mevc}{\ensuremath{\aunit{Me\kern -0.1em V\!/}c}\xspace}
\newcommand{\gevc}{\ensuremath{\aunit{Ge\kern -0.1em V\!/}c}\xspace}
\newcommand{\mevcc}{\ensuremath{\aunit{Me\kern -0.1em V\!/}c^2}\xspace}
\newcommand{\gevcc}{\ensuremath{\aunit{Ge\kern -0.1em V\!/}c^2}\xspace}
\def\fb   {\ensuremath{\aunit{fb}}\xspace}
\def\invfb   {\ensuremath{\fb^{-1}}\xspace}

\newcommand{\chisqip}{\ensuremath{\chi^2_{\text{IP}}}\xspace}

\def\gsim{{~\raise.15em\hbox{$>$}\kern-.85em
          \lower.35em\hbox{$\sim$}~}\xspace}
\def\lsim{{~\raise.15em\hbox{$<$}\kern-.85em
          \lower.35em\hbox{$\sim$}~}\xspace}

\def\pt         {\ensuremath{p_{\mathrm{T}}}\xspace}

\def\tell1  {TELL1\xspace}
\def\ukl1   {UKL1\xspace}

\newcommand{\lhcborcid}[1]{\href{https://orcid.org/#1}{\hspace*{0.1em}\raisebox{-0.45ex}{\includegraphics[width=1em]{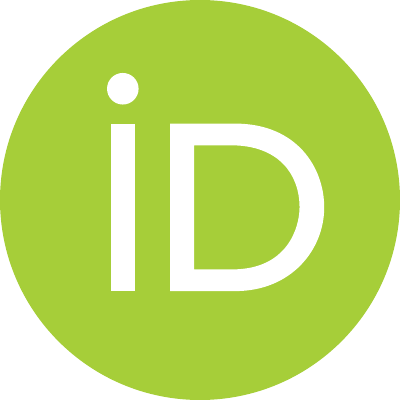}}}}

\usepackage{cite} % Allows for ranges in citations
\usepackage{mciteplus}
\usepackage{longtable} % only for template; not usually to be used in PAPERs
\begin{document}
\renewcommand{\thefootnote}{\fnsymbol{footnote}}
\setcounter{footnote}{1}
\begin{titlepage}
    \pagenumbering{roman}
    \vspace*{-1.5cm}
    \centerline{\large EUROPEAN ORGANIZATION FOR NUCLEAR RESEARCH (CERN)}
    \vspace*{1.5cm}
    \noindent
    \begin{tabular*}{\linewidth}{lc@{\extracolsep{\fill}}r@{\extracolsep{0pt}}}
    \ifthenelse{\boolean{pdflatex}}% Logo format choice
    {\vspace*{-1.5cm}\mbox{\!\!\!\includegraphics[width=.14\textwidth]{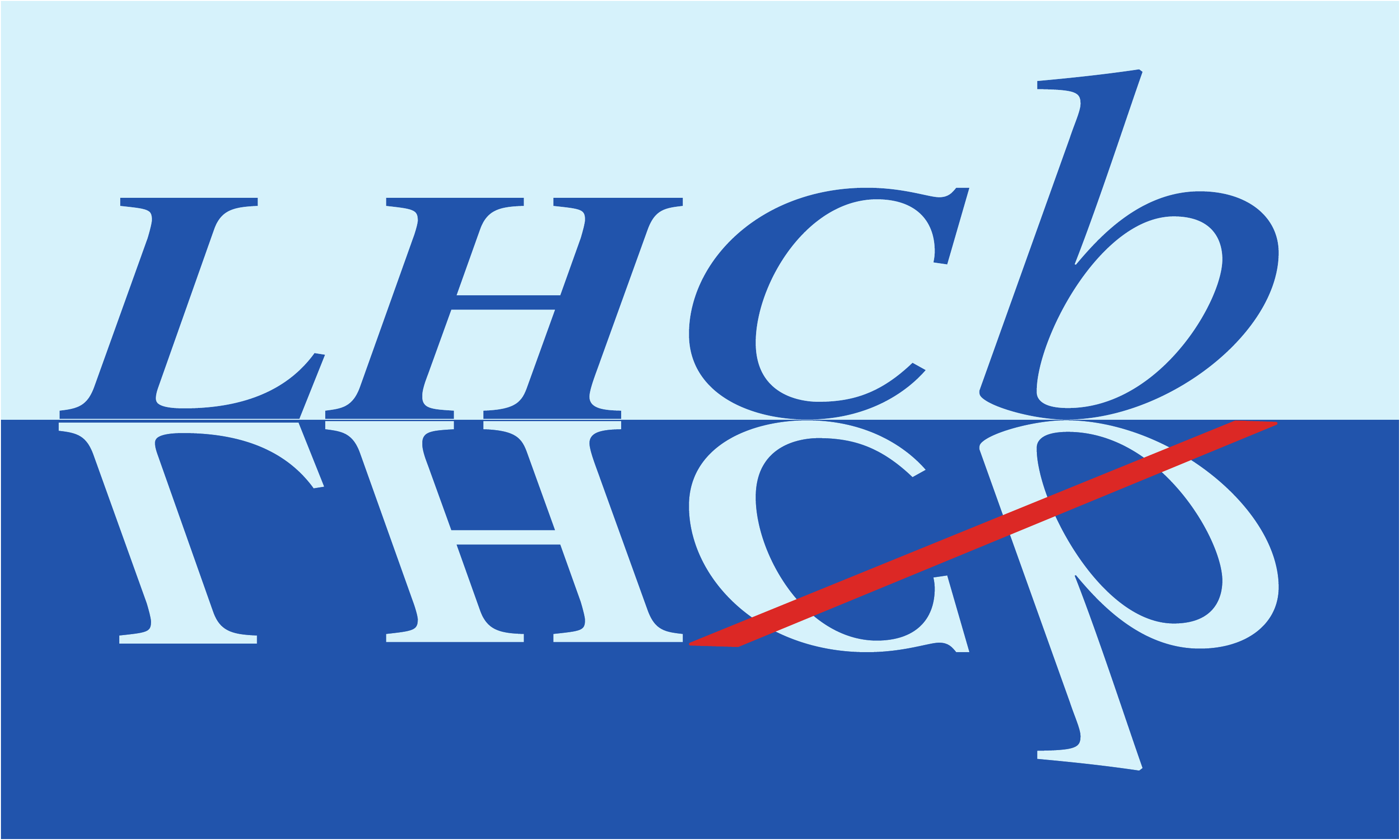}} & &}%
    {\vspace*{-1.2cm}\mbox{\!\!\!\includegraphics[width=.12\textwidth]{lhcb-logo.eps}} & &}%
    \\
     & & CERN-EP-2023-006 \\  % ID 
     & & LHCb-PAPER-2022-043 \\  % ID 
     & & September 27, 2023 \\ % Date - Can also hardwire e.g.: 23 March 2010
     & & \\
    \end{tabular*}
    \vspace*{4.0cm}
    {\normalfont\bfseries\boldmath\huge
    \begin{center}
    \papertitle 
    \end{center}
    }
    \vspace*{2.0cm}
    \begin{center}
    \paperauthors \footnote{Authors are listed at the end of this Letter.}
    \end{center}
    \vspace{\fill}
    \begin{abstract}
    \noindent
    Two new excited states, $\POmega_c(3185)^0$ and $\POmega_c(3327)^0$, are observed in the $\Xicp\Km$ invariant-mass spectrum using proton-proton collision data collected by the LHCb experiment, corresponding to an integrated luminosity of $9\invfb$. Five previously observed excited $\POmega_c^0$ states are confirmed, namely $\POmega_c(3000)^0$, $\POmega_c(3050)^0$, $\POmega_c(3065)^0$, $\POmega_c(3090)^0$, and $\POmega_c(3119)^0$. The masses and widths of these seven states are measured with the highest precision to date.\\
    \end{abstract}
    \vspace*{2.0cm}
    \begin{center}
      Published in Phys. Rev. Lett. 131 (2023) 131902
    \end{center}
    \vspace{\fill}
    {\footnotesize 
    \centerline{\copyright~\papercopyright. \href{\paperlicenceurl}{\paperlicence}.}}
    \vspace*{2mm}
    \end{titlepage}
    \newpage
    \setcounter{page}{2}
    \mbox{~}
    
\renewcommand{\thefootnote}{\arabic{footnote}}
\setcounter{footnote}{0}
\pagestyle{plain} % restore page numbers for the main text
\setcounter{page}{1}
\pagenumbering{arabic}
\newcommand{\Oc}{\POmega_c^0}
\newcommand{\OcX}[1]{\POmega_c(#1)^{0}}
\newcommand{\Ocstst}{\OcX{X}}

Singly charmed baryons consist of one charm quark and two lighter quarks. 
Due to the large mass difference between the charm quark and the lighter quarks, the singly charmed baryon mass spectrum can be described
systematically in heavy quark effective theory (HQET)~\cite{Grozin:1992yq}.
Many excited states are expected because of the complex interplay of the three-quark system, which makes them an ideal testing ground for theories of the strong force. Studies of charmed baryon spectroscopy have the potential to reveal important insights into the fundamental nature of hadronic matter.

In 2017, the $\lhcb$ collaboration reported the observation of five new narrow $\Oc$ states decaying to the $\PXi_c^+ K^-$ final state~\cite{LHCb-PAPER-2017-002}. Four of them have been confirmed in $e^+e^-$ collisions by the Belle collaboration~\cite{Yelton:2017qxg} and in the $\PXi_c^+ K^-$ invariant-mass projection of $\Omegab \to \Xicp K^- \pi^-$ decays by the $\lhcb$ collaboration~\cite{LHCb-PAPER-2021-012}. The inclusion of charge-conjugate processes is implied throughout this Letter, unless stated otherwise.

The $\Oc$ states have been studied extensively from the theoretical side. Most theoretical studies interpret these states as excited bound states~\cite{Karliner:2017kfm,Chen:2017sci,Wang:2017hej,Wang:2017vnc,Cheng:2017ove,Zhao:2017fov,Agaev:2017lip,Chen:2017gnu,Wang:2017zjw,Wang:2017kfr,Yao:2018jmc,Mao:2017wbz,Santopinto:2018ljf,Xu:2019kkt,Hu:2020nkg,Yang:2021lce,RamirezMorales:2022lsy}; however, the different schemes for matching their masses and quantum numbers have resulted in conflicting interpretations among different studies.
In some studies~\cite{Debastiani:2017ewu,Montana:2017kjw,Ramos:2020bgs}, the $\OcX{3050}$ and $\OcX{3090}$ states are interpreted as baryon-meson molecular (quasi-bound) states. Some theoretical studies~\cite{An:2017lwg,Wang_2021} also suggest the presence of three pentaquark states (\squark\squark\cquark\qqbar) with masses very close to the $\OcX{3065}$, $\OcX{3090}$, and $ \OcX{3119}$ states, which may indicate the presence of a considerable \squark\squark\cquark\qqbar component in these three states. In addition, another \squark\squark\cquark\qqbar state is predicted with a mass of $\sim$2980\mev (natural units are used throughout this Letter). The branching fractions of excited $\Oc$ baryons to final states such as $\PXi_c \pi$, $\POmega_c \eta$, etc., may depend on the internal molecular or pentaquark structure of these baryons, and thus provide a way to elucidate their inner structure. Futhermore, the search for new excited states is also an essential way to experimentally study the charmed baryon spectrum. A lattice QCD calculation~\cite{Padmanath:2017lng} presents a rich invariant-mass spectrum, dozens of D-- or F--wave excited states were identified, expanding our understanding of the charmed baryon spectrum. Studying the excited states of $\Oc$ baryons provides a unique opportunity to explore the properties of these intriguing particles, and to examine and verify a variety of theoretical predictions, including phenomenological models, thereby deepening our understanding of fundamental strong interaction effects.

This Letter presents an updated analysis of the $\PXi_c^+ K^-$ invariant-mass spectrum, using proton-proton ($pp$) collision data collected by the LHCb experiment from 2011 to 2018 at center-of-mass energies of 7, 8 and 13$\tev$, corresponding to integrated luminosities of $1$, $2$ and $6\invfb$, respectively. Grouped by trigger strategy, data recorded in 2011--2015 is referred to as data set 1, and data collected in 2016--2018 is referred to as data set 2. The increased charm production cross section at higher center-of-mass energy, the higher integrated luminosity, and the improvement in the trigger system~\cite{LHCb-DP-2012-004,LHCb-DP-2019-001} result in a five times larger sample size compared to the previous analysis~\cite{LHCb-PAPER-2017-002}. The results presented in this Letter supersede the previous mass and width determinations~\cite{LHCb-PAPER-2017-002}.

The $\lhcb$ detector~\cite{Alves:2008zz,LHCb-DP-2014-002} is a single-arm forward spectrometer covering the pseudo-rapidity range $2<\eta<5$, designed for the study of particles containing $\bquark$ or $\cquark$ quarks. 
Simulated data samples are produced with the software packages described in Refs.~\cite{Sjostrand:2007gs,LHCb:2011dpk,Lange:2001uf,Allison:2006ve,Clemencic:2011zza} and are used to develop the event selection, estimate the invariant-mass resolution, and model physics processes that may constitute peaking backgrounds in the analysis.

The $\PXi_c^+(\to \proton \Km \pip)$ candidates are formed with three tracks that do not originate from any primary $pp$ interaction vertex (PV) and have significant transverse momentum ($\pt$). Particle identification (PID) requirements are applied to all the final-state tracks in order to suppress combinatorial background. Additionally, PID vetoes 
are applied to suppress background from misidentified charm meson decays. These vetoes are 
formed by placing requirements on the invariant-mass of the candidates recomputed with the relevant change in the particle mass hypothesis. The $\PXi_c^+$ candidates are required to have a large flight-distance significance from any PV and a small value of $\chisqip$, defined as the difference between the vertex fit $\chi^2$ of the PV reconstructed with and without the particle in question.
The reconstructed $\PXi_c^+$ candidates must have an invariant mass in the range of $[2450,2485] \mev$, corresponding to about three times the invariant-mass resolution around the known $\Xicp$ mass~\cite{PDG2022}. Each $\PXi_c^+$ candidate is combined with a $K^-$ candidate that originates from the same PV and has $\pt$ greater than $0.4\gev$. 
The $\PXi_c^+K^-$ combinations
with good vertex quality and $\pt$ greater than $4.5\gev$ are retained as $\Ocstst$ candidates, where $X$ is the mass of the excited state. 

A multivariate classifier based on a boosted decision tree (BDT) algorithm~\cite{BBDT} and implemented in the TMVA~\cite{TMVA4} toolkit is used to further improve the signal purity. The variables used to train the BDT classifier are: the $\chi^2$ value of the $\PXi_c^+$ decay-vertex fit, the $\PXi_c^+$ flight distance, the angle between the $\PXi_c^+$ momentum vector and the line that connects the $\PXi_c^+$ decay vertex with its PV, the $\chisqip$ and $\pt$ of $\PXi_c^+$ and $\Ocstst$ candidates; and the $\chisqip$, $\pt$ and particle-identification variables of all final state particles. 
Due to the difference in trigger strategies, different BDT classifiers are used for data sets 1 and 2. The BDT classifier is trained with simulated signal decays generated with the 2012 (2016) configuration for data set 1 (data set 2), and background candidates taken from the $\PXi_c^+$ invariant-mass sideband region, defined as $[2400, 2440] \mev$ and $[2500, 2540] \mev$.
The working points are chosen such that the BDT classifier efficiency on the signal is 75\%; no fine-tuned optimization is performed to avoid favoring any particular excited state.

\begin{figure}
\begin{center}
    {\includegraphics[width=0.30\linewidth]{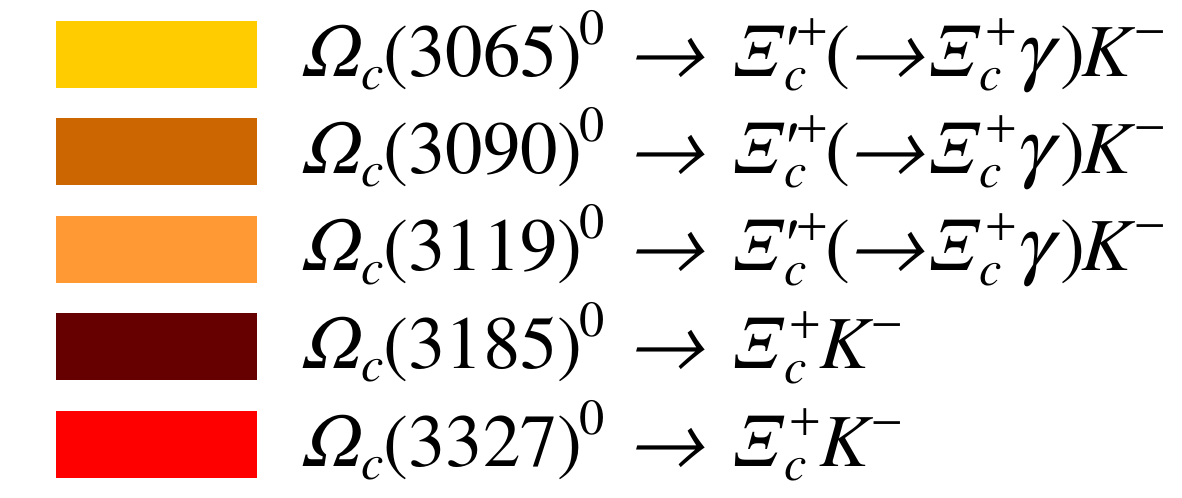}}
    {\includegraphics[width=0.30\linewidth]{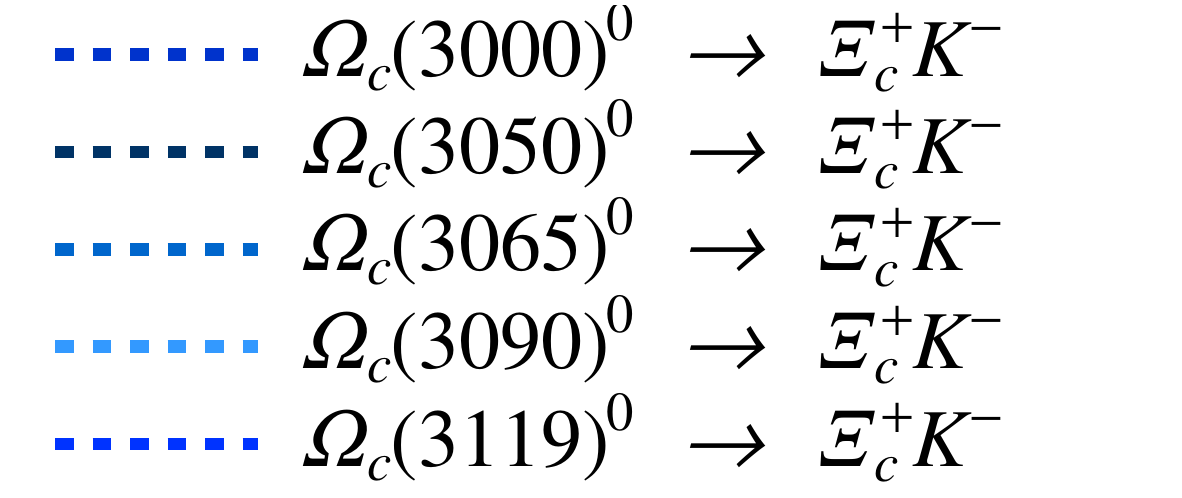}}
    {\includegraphics[width=0.30\linewidth]{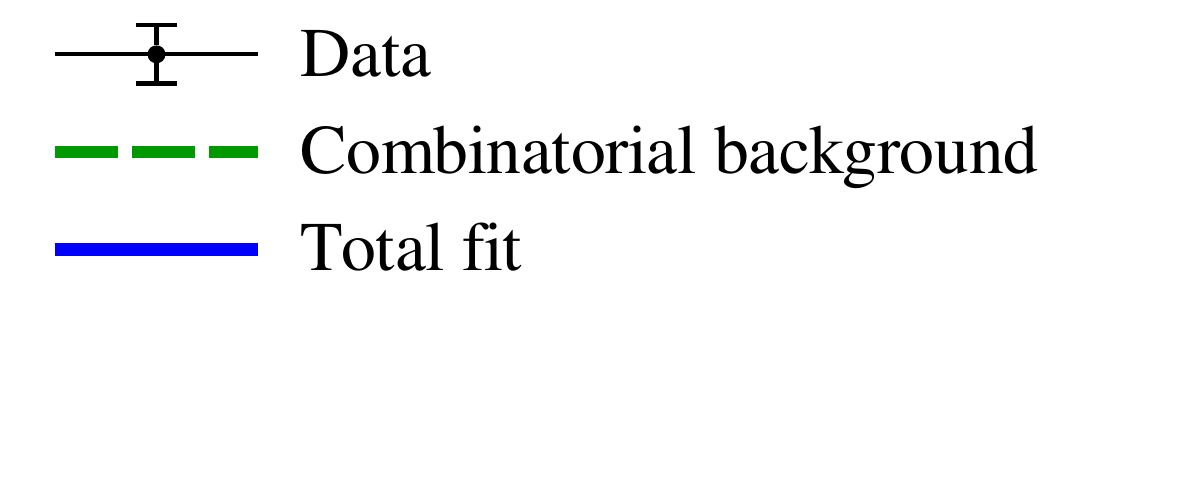}}
    \subfigure{\includegraphics[width=0.49\linewidth]{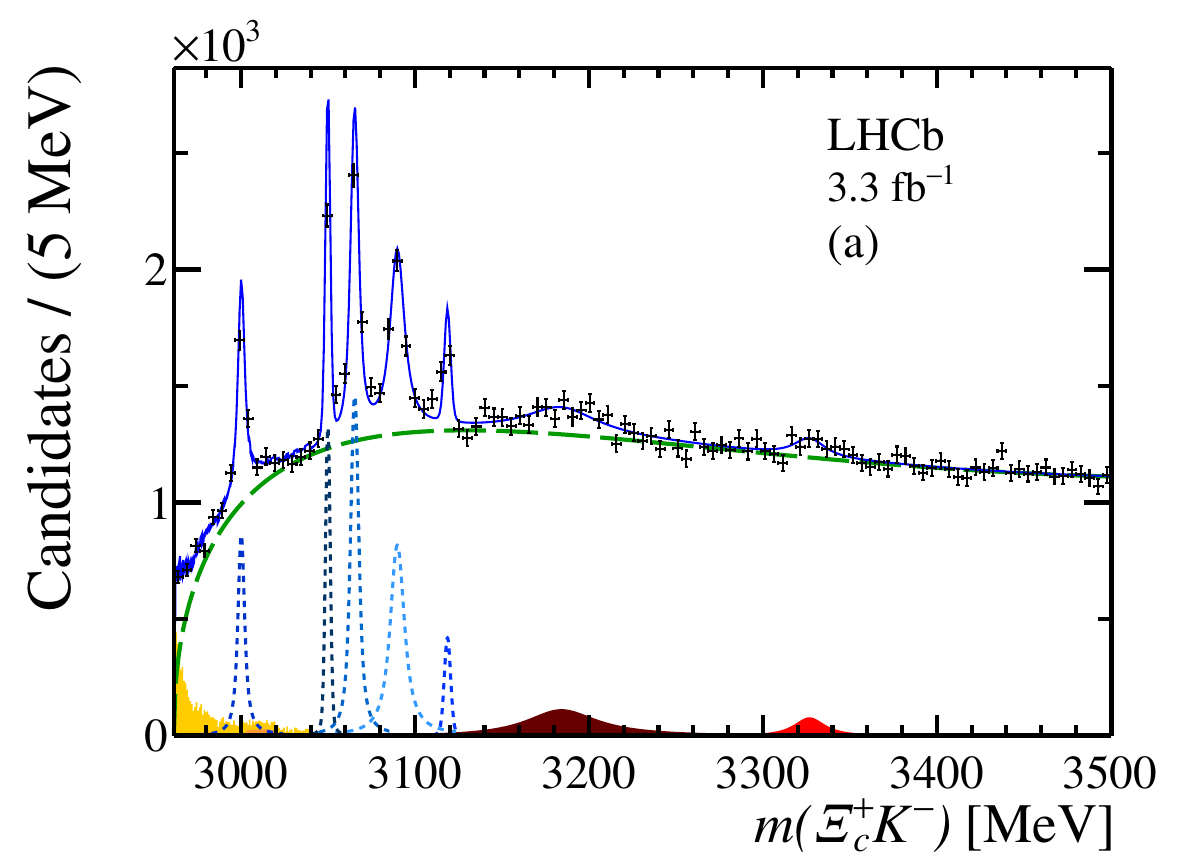}}
    \subfigure{\includegraphics[width=0.49\linewidth]{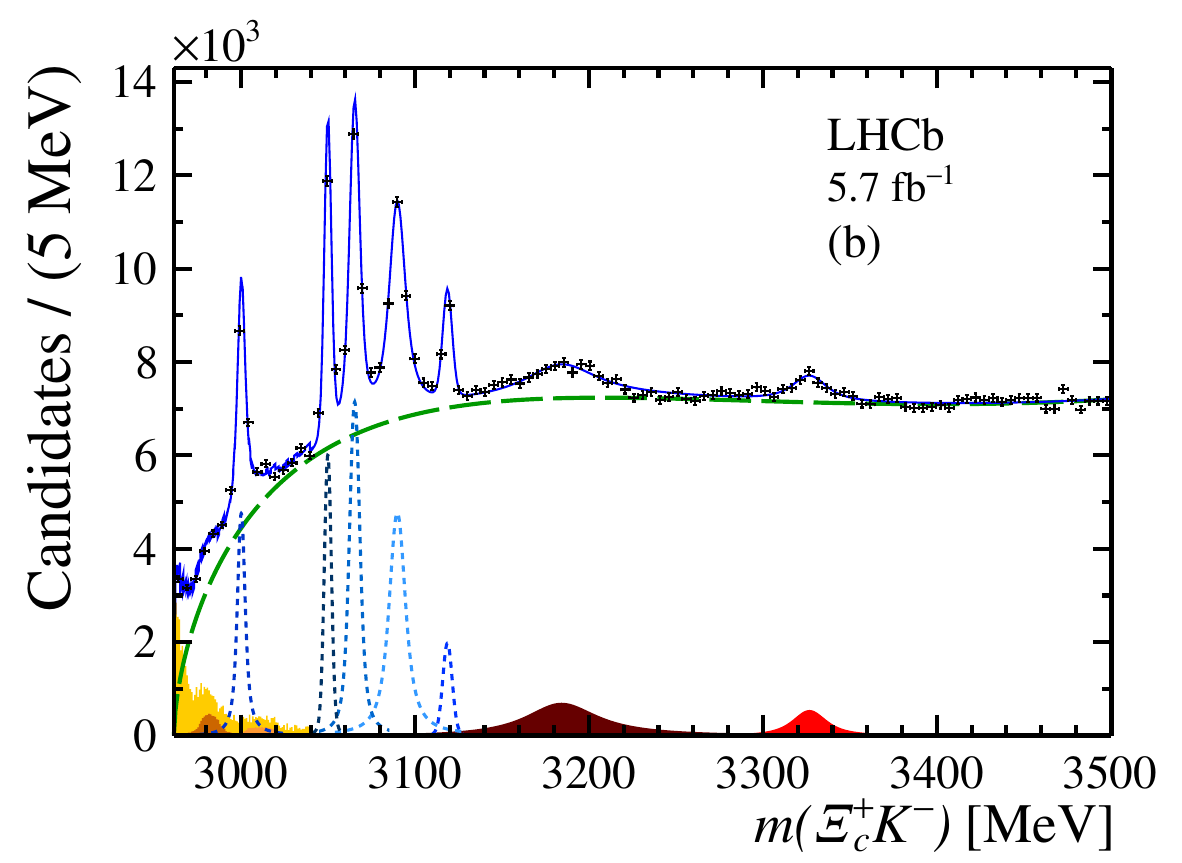}}
  \end{center}
\caption{\small
    Invariant-mass distribution of the $\Ocstst$ candidates in (a) data set 1 and (b) data set 2, with the fit results overlaid. A bin width of 5\mev is used for plotting. The previously observed excited $\Oc$ states are shown in blue dashed lines. The $\OcX{3185}$ state is shown in the brown area, and the $\OcX{3327}$ state is shown in the red area. Three feed-down components are shown as the yellow areas, while the green long-dashed line corresponds to the combinatorial background.}
  \label{fig:result}
\end{figure}

To improve the mass resolution, the variable $m(\Xicp K^-)$ is defined as the difference between the invariant mass of the $\Ocstst$ and $\Xicp$ candidates, to which the known $\Xicp$ mass~\cite{PDG2022} is added. 
The $\Xicp K^-$ invariant-mass distribution is shown in Fig.~\ref{fig:result}, where seven peaking structures are seen. Five of them have been observed in the previous analysis~\cite{LHCb-PAPER-2017-002}, while the $\OcX{3185}$ and $\OcX{3327}$ states are observed for the first time. There are no similar structures in the wrong-sign sample ($\Xicp K^+$) or in the spectrum of $\PXi_c^+$ sideband candidates combined with a kaon. 

To determine the masses and widths of these $\Ocstst$ states, an extended maximum likelihood fit with bin widths of 1$\mev$ is performed to the $\Xicp K^-$ invariant-mass distributions, simultaneously to data sets 1 and 2. 
The $\Ocstst$ contributions are described by S-wave relativistic Breit--Wigner functions convolved with a Gaussian resolution function whose width is determined from simulated signal samples. 
The combinatorial background is parameterized by an empirical function,
\begin{equation}
      {B}(\Delta m)=(\Delta m)^{a}\times \exp(b_1 \Delta m+b_2 \Delta m^2 ),\label{equ:BKG}
\end{equation}
where $\Delta m$ is the difference between the invariant mass of the $\Xicp K^-$ candidate and the $\PXi_c^+K^-$ mass threshold, and $a$, $b_1$ and $b_2$ are free parameters.
In addition, three feed-down components from partially reconstructed decays of the $\OcX{3065}$, $\OcX{3090}$, and $\OcX{3119}$ resonances are included. These contributions are from $\OcX{X} \to \PXi_c^{\prime +}(\to \PXi_c^+ \gamma) K^-$ decays, where the photon is not reconstructed. Their shapes are determined from simulated samples and are fixed, while their yields are free to vary in the fit. 

\begin{table}[b]
\caption{\small Fit results of the mass, width, and yield for each state, and for each data set. Uncertainties are statistical only.}\label{tab:result}
\begin{center}
\begin{tabular}{ccccc}
\hline
Resonance           & $m$ (MeV)                           & $\Gamma$ (MeV) & Yield (data set 1) & Yield (data set 2) \\
\hline
$\OcX{3000}$  & $3000.44\pm0.07$ &$\phantom{0}3.83\pm0.23$   &$1225\pm83\phantom{0}$    &$\phantom{0}7533\pm263\phantom{0}$  \\
$\OcX{3050}$  & $3050.18\pm0.04$ &$\phantom{0}0.67\pm0.17$   &$1139\pm65\phantom{0}$   &$\phantom{0}7379\pm215\phantom{0}$  \\
$\OcX{3065}$  & $3065.63\pm0.06$ &$\phantom{0}3.79\pm0.20$   &$2180\pm99\phantom{0}$               &$13046\pm316\phantom{0}$  \\
$\OcX{3090}$  & $3090.16\pm0.11$ &$\phantom{0}8.48\pm0.44$   &$2234\pm136$               &$14434\pm486\phantom{0}$  \\
$\OcX{3119}$  & $3118.98\pm0.12$ &$\phantom{0}0.60\pm0.63$   &$\phantom{0}470\pm66\phantom{0}$     &$\phantom{0}3279\pm234\phantom{0}$  \\
$\OcX{3185}$  & $3185.1\pm1.7$ &$50\pm7$          &$1642\pm367$               &$10278\pm1565$  \\
$\OcX{3327}$  & $3327.1\pm1.2$ &$20\pm5$             &$\phantom{0}489\pm173$     &$\phantom{0}3649\pm723\phantom{0}$ \\
\hline
\end{tabular}
\end{center}
\end{table}

The fit results are shown in Fig.~\ref{fig:result}
and summarized in Table~\ref{tab:result}. 
In addition to 
the five narrow states at lower invariant mass that were previously observed, two new states with masses of about 3185\mev and 3327\mev, denoted $\OcX{3185}$ and $\OcX{3327}$ are observed with high significance.

The enhancement around the $\Xicp K^-$ mass threshold is described by the partially reconstructed decays of $\OcX{X} \to \PXi_c^{\prime +}(\to \PXi_c^+ \gamma) K^-$, as was done in the previous analysis~\cite{LHCb-PAPER-2017-002}. In the exclusive analysis using the \mbox{$\POmega_b^-\to\Ocstst(\to\PXi_c^+K^-)\pi^-$} decay~\cite{LHCb-PAPER-2021-012}, these feed-down components are excluded by requiring an appropriate signal mass window of the $\POmega_b^-$ baryon, while the threshold enhancement is still present. This enhancement was modeled as an S-wave Breit--Wigner distribution, but the available data in Ref.~\cite{LHCb-PAPER-2021-012} are not sufficient to declare an observation. 
To check if this resonance structure is present in this analysis, the enhancement is fitted using two alternative models: a Breit--Wigner component with and without the feed down coming from the $\OcX{3065}$ contribution. The yield of the $\OcX{3065}$ feed down is constrained to be 10\% of the $\OcX{3065}$ signal yield for the former. This constraint is equivalent to fixing the ratio $\mathcal{B}(\OcX{3065} \to \PXi_c^{\prime +} K^-)\times\mathcal{B}(\PXi_c^{\prime +} \to \PXi_c^{+}\gamma)/\mathcal{B}(\OcX{3065} \to \PXi_c^{+} K^-)$ to 0.1, which is chosen due to the small phase space of the $\OcX{3065} \to \PXi_c^{\prime +} K^-$ decay.
The yields of the other feed-down contributions are free to float. 
Unfortunately, the shape of the $\OcX{3065}$ feed down and the additional Breit--Wigner structure are too similar to be separated in this analysis. The relative contributions from these two components cannot be determined from data, and hence this is accounted for as a systematic uncertainty. The existence of another hidden state cannot be excluded.

The mass difference between the $\OcX{3185}$ and $\OcX{3327}$ baryons is approximately the mass of the pion. It is possible that partially reconstructed candidates from the $\OcX{3327}$ decay fall into the $\OcX{3185}$ region, such as \mbox{$\OcX{3327} \to \PXi_c(2645)^{+}(\to \PXi_c^+ \pi^0) K^-$} and $\OcX{3327} \to \PXi_c^{\prime +}(\to \PXi_c^+ \gamma) K^-$ decays with the $\pi^0$ or $\gamma$ not reconstructed. The $m(\PXi_c^+ K^-)$ line shapes of such contributions are obtained using the fast simulation toolkit RapidSim~\cite{Cowan:2016tnm}. The feed-down contributions have been studied under different spin structure hypotheses, and differences from phase-space simulation were found to be negligible.
The fit favors the presence of the $\OcX{3327}$ feed down with the $\piz$ missing. The $\OcX{3327}$ feed down strongly affects the measured mass and width of the $\OcX{3185}$ baryon, and is assigned as a systematic uncertainty. None of the models for the $\OcX{3185}$ region can be ruled out.
In addition, the effects of the three-body decay $\OcX{3327} \to \PXi_c^+ K^- \pi^0$ are checked with a uniform phase-space model, and found to be insignificant on any of the states when assuming the ratio $\mathcal{B}(\OcX{3327} \to \PXi_c^+ K^- \pi^0)/\mathcal{B}(\OcX{3327} \to \PXi_c^+ K^-)$ to be unity.

The effects of the \mbox{$\PXi_c(3055)^0 \to \PXi_{c}^{\prime+} (\to \PXi_c^+ \gamma) \pi^- $},
\mbox{$\PXi_c(3055)^0 \to \PXi_c(2645)^+ (\to \PXi_c^+ \piz) \pi^-$}, and \mbox{$\PXi_c(2923/2939/2965)^0 \to \PXi_c^+ \pi^-$} decays, for which the pion is misidentified as a kaon and the $\piz/\gamma$ from the $\PXi_c(2645)^+/\PXi_{c}^{\prime+}$ decays is missing, are studied and found to be negligible. The contribution from $\OcX{3327} \to \PXi_c^+ K^{*-}$ is kinematically suppressed.

Several checks are performed to confirm the existence of the observed states and the stability of the fitted parameters. Each data set is divided into subsamples according to data-taking conditions, charge combination ($\PXi_c^+K^-$ or $\PXi_c^-K^+$), or different intervals of  $\pt(\Km)$ and $\pt(\PXi_c^+)$. In all tests, the results are consistent with the default fit. 
A two-peak structure also describes the data well in the mass region around 3185\mev, hence the presence of two states in this region cannot be excluded. 
The bias from the fit model itself is estimated using pseudoexperiments. For each parameter, the mean differences between the fitted values and the input ones are included as systematic uncertainties.

\begin{sidewaystable}
\caption{\small Systematic uncertainties on the measured masses and natural widths.}\label{tab:sys}
\begin{center}
\resizebox{\textwidth}{!}{
\begin{tabular}{c|ccccccc}
\hline
Source & $\OcX{3000}$ & $\OcX{3050}$ & $\OcX{3065}$ & $\OcX{3090}$ & $\OcX{3119}$ & $\OcX{3185}$ & $\OcX{3327}$ \\ 
$\left[\mev,~\mev \right]$  & [$m,~\Gamma$] & [$m,~\Gamma$] & [$m,~\Gamma$] & [$m,~\Gamma$] & [$m,~\Gamma$] & [$m,~\Gamma$] & [$m,~\Gamma$] \\
\hline 
Threshold structure        &\makecell*[c]{\large [$^{+0.01}_{-0.07},~^{+1.23}_{-0.19}$]}  &\large [$^{+0.00}_{-0.01},~^{+0.00}_{-0.17}$]  &\large [$^{+0.01}_{-0.00},~^{+0.00}_{-0.11}$] &\large [$^{+0.01}_{-0.00},~^{+0.12}_{-0.20}$]  &\large [$^{+0.06}_{-0.00},~^{+0.00}_{-0.16}$]  &\large [$^{+0.32}_{-0.57},~^{+8.05}_{-0.83}$]  &\large [$^{+0.01}_{-0.15},~^{+1.71}_{-0.04}$]  \\[8pt]
feed down from $\Omegac(3327)$       &\large [$^{+0.04}_{-0.00},~^{+0.04}_{-0.00}$]  &\large [$^{+0.00}_{-0.00},~^{+0.00}_{-0.01}$]  &\large [$^{+0.01}_{-0.00},~^{+0.03}_{-0.00}$]  &\large [$^{+0.01}_{-0.00},~^{+0.22}_{-0.00}$]  &\large [$^{+0.01}_{-0.00},~^{+0.07}_{-0.00}$] &\large [$^{+4.28}_{-0.00},~^{+0.00}_{-10.81}$]  &\large [$^{+0.00}_{-0.20},~^{+0.98}_{-0.00}$]  \\[8pt]
Fit bias    &\large 
[$^{+0.02}_{-0.00},~^{+0.01}_{-0.00}$] &\large [$^{+0.02}_{-0.00},~^{+0.12}_{-0.00}$]  &\large [$^{+0.00}_{-0.00},~^{+0.08}_{-0.00}$]  &\large [$^{+0.01}_{-0.00},~^{+0.20}_{-0.00}$]  &\large [$^{+0.00}_{-0.02},~^{+0.40}_{-0.00}$]  &\large [$^{+0.00}_{-0.37},~^{+4.84}_{-0.00}$]  &\large [$^{+0.00}_{-0.18},~^{+0.00}_{-0.70}$]  \\[8pt] 
Spin and radius                     &\large [$^{+0.01}_{-0.01},~^{+0.01}_{-0.01}$]  &\large [$^{+0.01}_{-0.01},~^{+0.01}_{-0.01}$]  &\large [$^{+0.01}_{-0.01},~^{+0.02}_{-0.01}$] &\large [$^{+0.01}_{-0.01},~^{+0.05}_{-0.01}$]  &\large [$^{+0.01}_{-0.01},~^{+0.06}_{-0.01}$]  &\large [$^{+0.01}_{-0.18},~^{+0.56}_{-0.01}$] &\large [$^{+0.01}_{-0.02},~^{+0.01}_{-0.60}$]  \\[8pt] 
Background                     &\large [$^{+0.04}_{-0.00},~^{+0.98}_{-0.00}$]  &\large [$^{+0.00}_{-0.01},~^{+0.00}_{-0.18}$]  &\large [$^{+0.00}_{-0.00},~^{+0.00}_{-0.29}$]  &\large [$^{+0.00}_{-0.06},~^{+0.00}_{-1.58}$]  &\large [$^{+0.00}_{-0.21},~^{+0.00}_{-0.60}$]  &\large [$^{+5.98}_{-0.00},~^{+0.00}_{-16.2}$]  &\large [$^{+0.00}_{-1.24},~^{+12.9}_{-0.00}$]  \\  [8pt]
Bin size      &\large [$^{+0.00}_{-0.10},~^{+0.06}_{-0.04}$]  &\large [$^{+0.00}_{-0.01},~^{+0.40}_{-0.03}$]  &\large [$^{+0.02}_{-0.03},~^{+0.08}_{-0.07}$]  &\large [$^{+0.01}_{-0.05},~^{+0.37}_{-0.09}$]  &\large [$^{+0.00}_{-0.06},~^{+0.00}_{-0.60}$]  &\large [$^{+0.14}_{-0.48},~^{+2.75}_{-0.50}$]  &\large [$^{+0.07}_{-0.00},~^{+1.16}_{-0.00}$]  \\[8pt]
Momentum calibration                      & [$\pm$0.01,~\phantom{0}-\phantom{0}]  & [$\pm$0.03,~\phantom{0}-\phantom{0}]  & [$\pm$0.03,~\phantom{0}-\phantom{0}]  & [$\pm$0.04,~\phantom{0}-\phantom{0}]  & [$\pm$0.05,~\phantom{0}-\phantom{0}]  & [$\pm$0.07,~\phantom{0}-\phantom{0}]  & [$\pm$0.11,~\phantom{0}-\phantom{0}]  \\[8pt]
Energy loss                         & [$\pm$0.04,~\phantom{0}-\phantom{0}]  & [$\pm$0.04,~\phantom{0}-\phantom{0}]  & [$\pm$0.04,~\phantom{0}-\phantom{0}]  & [$\pm$0.04,~\phantom{0}-\phantom{0}]  & [$\pm$0.04,~\phantom{0}-\phantom{0}]  & [$\pm$0.04,~\phantom{0}-\phantom{0}]  & [$\pm$0.04,~\phantom{0}-\phantom{0}]  \\[4pt]
Simulation/Data difference     &\large [$^{+0.00}_{-0.01},~^{+0.20}_{-0.22}$]  &\large [$^{+0.01}_{-0.04},~^{+0.48}_{-0.67}$]  &\large [$^{+0.01}_{-0.00},~^{+0.36}_{-0.35}$]  &\large [$^{+0.00}_{-0.02},~^{+0.36}_{-0.29}$]  &\large [$^{+0.03}_{-0.00},~^{+0.80}_{-0.59}$]  &\large [$^{+0.08}_{-0.34},~^{+2.58}_{-0.67}$]  &\large [$^{+0.02}_{-0.05},~^{+0.27}_{-0.28}$]  \\[8pt]
\hline
Total                               & \makecell*[c]{\large [$^{+0.07}_{-0.13},^{+1.59}_{-0.29}$]}  & \large [$^{+0.06}_{-0.07},^{+0.64}_{-0.72}$]  & \large [$^{+0.06}_{-0.06},^{+0.38}_{-0.47}$]  & \large [$^{+0.06}_{-0.10},^{+0.61}_{-1.62}$]  & \large [$^{+0.09}_{-0.23},^{+0.90}_{-1.05}$]  & \large [$^{+7.36}_{-0.92},^{+10.1}_{-19.5}$]  & \large [$^{+0.14}_{-1.28},^{+13.1}_{-0.96}$]  \\[6pt]
\hline
\end{tabular}}
\end{center}
\end{sidewaystable}

The uncertainty related to the signal model is estimated by fitting the data with variations in the spin hypotheses and the Blatt-Weisskopf factor~\cite{Blatt:1952ije}.
The systematic uncertainty from the combinatorial background model is estimated by using a fourth-order Chebyshev polynomial as an alternative function. In addition, a systematic uncertainty is assigned based on the spread of results obtained by performing the fit with different bin widths.

Some sources of systematic uncertainty only contribute to the mass measurement, including the momentum calibration for charged particles, which has a relative accuracy of 0.03\%~\cite{LHCb-PAPER-2012-048,LHCb-PAPER-2013-011}, and the energy loss due to the imperfect modeling of the detector material, which results in a $0.04\mev$ uncertainty. The differences between simulation and data mainly affect the measurement of the natural width of each state. It is calculated by varying the width of the resolution function by $10\%$~\cite{LHCB-PAPER-2014-048}. Other assumptions for the $\OcX{3185}$ structure, such as the two-peak structure, 
additional feed-down components or a combination of the two-peak structure and one more feed-down assumption cannot be excluded, and  
are included in the systematic uncertainties.

The systematic uncertainties are summarized in Table~\ref{tab:sys}. 
After considering the systematic uncertainty, the significances of the $\OcX{3185}$ and $\OcX{3327}$ states are larger than 12\,$\sigma$ and 10\,$\sigma$, respectively. The significance for each state is determined from Wilk's theorem using the difference in log-likelihood with and without that signal component. 
% Other assumptions, such as the two-peak structure, 
% additional feed-down components or a combination of them cannot be excluded, and  
% are included in the systematic uncertainties.
The results of the measured mass and natural width of all seven states 
, shown in Table~\ref{tab:finalresult}, are the most precise to date, and supersede those in Ref.~\cite{LHCb-PAPER-2017-002}. The correlations between 
these results and those in  
Ref.~\cite{LHCb-PAPER-2021-012} are negligible, 
except for the systematic uncertainty due to the momentum calibration and energy loss, which are fully correlated. 
The natural widths of the $\OcX{3050}$ and $\OcX{3119}$ baryons are very close to zero; therefore upper limits on them are set at Bayesian 95\% confidence level, assuming Gaussian behavior for both statistical and systematic uncertainties.

\begin{table}
\caption{\small Measured mass and natural width for each of the seven $\OcX{X}$ states. The first uncertainty is statistical and the second is systematic, the third (mass only) arises from the uncertainty of the known $\PXi_{c}^{+}$ mass.}\label{tab:finalresult}
\begin{center}
\begin{tabular}{ccc}
\hline
Resonance           & $m$ (MeV)                           & $\Gamma$ (MeV) \\
\hline 
$\noindent\rule{0pt}{0.9\normalbaselineskip}\OcX{3000}$  & $3000.44\pm0.07^{\phantom{0}+0.07}_{\phantom{0}-0.13}\pm0.23$ &$\phantom{0}3.83\pm0.23^{\phantom{0}+1.59}_{\phantom{0}-0.29} $    \\
$\OcX{3050}$  & $3050.18\pm0.04^{\phantom{0}+0.06}_{\phantom{0}-0.07}\pm0.23$ &$\phantom{0}0.67\pm0.17^{\phantom{0}+0.64}_{\phantom{0}-0.72} $    \\
& &\small$\phantom{==}<1.8\mev, 95\%\ {\rm C.L.}$ \\
$\OcX{3065}$  & $3065.63\pm0.06^{\phantom{0}+0.06}_{\phantom{0}-0.06}\pm0.23$ &$\phantom{0}3.79\pm0.20^{\phantom{0}+0.38}_{\phantom{0}-0.47} $    \\
$\OcX{3090}$  & $3090.16\pm0.11^{\phantom{0}+0.06}_{\phantom{0}-0.10}\pm0.23$ &$\phantom{0}8.48\pm0.44^{\phantom{0}+0.61}_{\phantom{0}-1.62} $    \\
$\OcX{3119}$  & $3118.98\pm0.12^{\phantom{0}+0.09}_{\phantom{0}-0.23}\pm0.23$ &$\phantom{0}0.60\pm0.63^{\phantom{0}+0.90}_{\phantom{0}-1.05} $    \\
& & \small$\phantom{==}<2.5\mev, 95\%\ {\rm C.L.}$ \\
$\OcX{3185}$  & $3185.1\pm1.7^{\phantom{0}+7.4}_{\phantom{0}-0.9}\pm0.2$ &$50\pm7^{\phantom{0}+10}_{\phantom{0}-20}$    \\
$\OcX{3327}$  & $3327.1\pm1.2^{\phantom{0}+0.1}_{\phantom{0}-1.3}\pm0.2$ &$20\pm5^{\phantom{0}+13}_{\phantom{0}-1}$    \\
\hline
\end{tabular}
\end{center}
\end{table}

In conclusion, the $\PXi_c^+ K^-$ invariant-mass spectrum is investigated using $pp$ collision data collected by the LHCb experiment. 
With a total integrated luminosity of $9\invfb$, we were able to select a high-purity sample of $\PXi_c^+$ candidates and observe seven excited states, including two that had never been seen before: the $\OcX{3185}$ and $\OcX{3327}$ states, and also the $\OcX{3119}$ state that is not seen in the exclusive analysis using the \mbox{$\POmega_b^-\to\Ocstst(\to\PXi_c^+K^-)\pi^-$} decay~\cite{LHCb-PAPER-2021-012}. The $\OcX{3185}$ and $\OcX{3327}$ states have masses close to the threshold of the $\PXi D$ and $\PXi D^*$ final states. 
Referring to the lattice QCD result of Ref.~\cite{Padmanath:2017lng}, the $\OcX{3185}$ mass is in the predicted range for P-wave states, and the $\OcX{3327}$ mass is in the range for many possible states. While the quantum numbers of these states remain to be determined, their observations 
provide new information on the complex hadron spectroscopy in order to develop a deeper understanding of the strong interaction and its underlying principles.

\section*{Acknowledgements}
%
% These Acknowledgements valid from 3-May-2019
%
\noindent We express our gratitude to our colleagues in the CERN
accelerator departments for the excellent performance of the LHC. We
thank the technical and administrative staff at the LHCb
institutes.
We acknowledge support from CERN and from the national agencies:
CAPES, CNPq, FAPERJ and FINEP (Brazil); 
MOST and NSFC (China); 
CNRS/IN2P3 (France); 
BMBF, DFG and MPG (Germany); 
INFN (Italy); 
NWO (Netherlands); 
MNiSW and NCN (Poland); 
MEN/IFA (Romania); 
%MSHE (Russia); 
MICINN (Spain); 
SNSF and SER (Switzerland); 
NASU (Ukraine); 
STFC (United Kingdom); 
DOE NP and NSF (USA).
We acknowledge the computing resources that are provided by CERN, IN2P3
(France), KIT and DESY (Germany), INFN (Italy), SURF (Netherlands),
PIC (Spain), GridPP (United Kingdom), 
%RRCKI and Yandex LLC (Russia), 
CSCS (Switzerland), IFIN-HH (Romania), CBPF (Brazil),
Polish WLCG  (Poland) and NERSC (USA).
We are indebted to the communities behind the multiple open-source
software packages on which we depend.
Individual groups or members have received support from
ARC and ARDC (Australia);
Minciencias (Colombia);
AvH Foundation (Germany);
EPLANET, Marie Sk\l{}odowska-Curie Actions and ERC (European Union);
A*MIDEX, ANR, IPhU and Labex P2IO, and R\'{e}gion Auvergne-Rh\^{o}ne-Alpes (France);
Key Research Program of Frontier Sciences of CAS, CAS PIFI, CAS CCEPP, 
Fundamental Research Funds for the Central Universities, 
and Sci. \& Tech. Program of Guangzhou (China);
%Key Research Program of Frontier Sciences of CAS, CAS PIFI,
%Thousand Talents Program, and Sci. \& Tech. Program of Guangzhou (China);
%RFBR, RSF and Yandex LLC (Russia);
GVA, XuntaGal, GENCAT and Prog.~Atracci\'on Talento, CM (Spain);
SRC (Sweden);
the Leverhulme Trust, the Royal Society
 and UKRI (United Kingdom).

\addcontentsline{toc}{section}{References}
% \setboolean{inbibliography}{false}

\ifx\mcitethebibliography\mciteundefinedmacro
\PackageError{LHCb.bst}{mciteplus.sty has not been loaded}
{This bibstyle requires the use of the mciteplus package.}\fi
\providecommand{\href}[2]{#2}

\newpage
% LHCb collaboration author list
% Data extracted on June 13th, 2023 at 4:06pm for paper reference LHCb-PAPER-2022-043
\centerline
{\large\bf LHCb collaboration}
\begin
{flushleft}
\small
R.~Aaij$^{32}$\lhcborcid{0000-0003-0533-1952},
A.S.W.~Abdelmotteleb$^{50}$\lhcborcid{0000-0001-7905-0542},
C.~Abellan~Beteta$^{44}$,
F.~Abudin{\'e}n$^{50}$\lhcborcid{0000-0002-6737-3528},
T.~Ackernley$^{54}$\lhcborcid{0000-0002-5951-3498},
B.~Adeva$^{40}$\lhcborcid{0000-0001-9756-3712},
M.~Adinolfi$^{48}$\lhcborcid{0000-0002-1326-1264},
P.~Adlarson$^{77}$\lhcborcid{0000-0001-6280-3851},
H.~Afsharnia$^{9}$,
C.~Agapopoulou$^{13}$\lhcborcid{0000-0002-2368-0147},
C.A.~Aidala$^{78}$\lhcborcid{0000-0001-9540-4988}
% ...

{\footnotesize \it

% ...
$^{9}$Universit{\'e} Clermont Auvergne, CNRS/IN2P3, LPC, Clermont-Ferrand, France\\
% ...
$^{13}$LPNHE, Sorbonne Universit{\'e}, Paris Diderot Sorbonne Paris Cit{\'e}, CNRS/IN2P3, Paris, France\\
% ...
$^{32}$Nikhef National Institute for Subatomic Physics, Amsterdam, Netherlands\\
% ...
$^{40}$Instituto Galego de F{\'\i}sica de Altas Enerx{\'\i}as (IGFAE), Universidade de Santiago de Compostela, Santiago de Compostela, Spain\\
% ...
$^{44}$Physik-Institut, Universit{\"a}t Z{\"u}rich, Z{\"u}rich, Switzerland\\
% ...
$^{48}$H.H. Wills Physics Laboratory, University of Bristol, Bristol, United Kingdom\\
% ...
$^{50}$Department of Physics, University of Warwick, Coventry, United Kingdom\\
% ...
$^{53}$School of Physics and Astronomy, University of Glasgow, Glasgow, United Kingdom\\
$^{54}$Oliver Lodge Laboratory, University of Liverpool, Liverpool, United Kingdom\\
% ...
$^{62}$Syracuse University, Syracuse, NY, United States\\
% ...
$^{77}$Department of Physics and Astronomy, Uppsala University, Uppsala, Sweden, associated to $^{53}$\\
$^{78}$University of Michigan, Ann Arbor, MI, United States, associated to $^{62}$\\
\bigskip
% ...
}
\end{flushleft}


\begin{mcitethebibliography}{10}
\mciteSetBstSublistMode{n}
\mciteSetBstMaxWidthForm{subitem}{\alph{mcitesubitemcount})}
\mciteSetBstSublistLabelBeginEnd{\mcitemaxwidthsubitemform\space}
{\relax}{\relax}

\bibitem{Grozin:1992yq}
A.~G. Grozin, \ifthenelse{\boolean{articletitles}}{\emph{{Introduction to the
  heavy quark effective theory. part 1}},
  }{}\href{http://arxiv.org/abs/hep-ph/9908366}{{\normalfont\ttfamily
  arXiv:hep-ph/9908366}}\relax
\mciteBstWouldAddEndPuncttrue
\mciteSetBstMidEndSepPunct{\mcitedefaultmidpunct}
{\mcitedefaultendpunct}{\mcitedefaultseppunct}\relax
\EndOfBibitem
\bibitem{LHCb-PAPER-2017-002}
LHCb collaboration, R.~Aaij {\em et~al.},
  \ifthenelse{\boolean{articletitles}}{\emph{{Observation of five new narrow
  $\Omegac$ states decaying to $\Xicp\Km$}},
  }{}\href{https://doi.org/10.1103/PhysRevLett.118.182001}{Phys.\ Rev.\ Lett.\
  \textbf{118} (2017) 182001},
  \href{http://arxiv.org/abs/1703.04639}{{\normalfont\ttfamily
  arXiv:1703.04639}}\relax
\mciteBstWouldAddEndPuncttrue
\mciteSetBstMidEndSepPunct{\mcitedefaultmidpunct}
{\mcitedefaultendpunct}{\mcitedefaultseppunct}\relax
\EndOfBibitem
\bibitem{Yelton:2017qxg}
Belle collaboration, J.~Yelton {\em et~al.},
  \ifthenelse{\boolean{articletitles}}{\emph{{Observation of excited
  $\Omegares_c$ charmed baryons in $e^+e^-$ collisions}},
  }{}\href{https://doi.org/10.1103/PhysRevD.97.051102}{Phys.\ Rev.\
  \textbf{D97} (2018) 051102},
  \href{http://arxiv.org/abs/1711.07927}{{\normalfont\ttfamily
  arXiv:1711.07927}}\relax
\mciteBstWouldAddEndPuncttrue
\mciteSetBstMidEndSepPunct{\mcitedefaultmidpunct}
{\mcitedefaultendpunct}{\mcitedefaultseppunct}\relax
\EndOfBibitem
\bibitem{LHCb-PAPER-2021-012}
LHCb collaboration, R.~Aaij {\em et~al.},
  \ifthenelse{\boolean{articletitles}}{\emph{{Observation of excited $\Omegac$
  baryons in \mbox{$\Omegab \to \Xic^+ \Km \pip$} decays}},
  }{}\href{https://doi.org/10.1103/PhysRevD.104.L091102}{Phys.\ Rev.\
  \textbf{D104} (2021) L091102},
  \href{http://arxiv.org/abs/2107.03419}{{\normalfont\ttfamily
  arXiv:2107.03419}}\relax
\mciteBstWouldAddEndPuncttrue
\mciteSetBstMidEndSepPunct{\mcitedefaultmidpunct}
{\mcitedefaultendpunct}{\mcitedefaultseppunct}\relax
\EndOfBibitem
\bibitem{Karliner:2017kfm}
M.~Karliner and J.~L. Rosner, \ifthenelse{\boolean{articletitles}}{\emph{{Very
  narrow excited $\Omegares_c$ baryons}},
  }{}\href{https://doi.org/10.1103/PhysRevD.95.114012}{Phys.\ Rev.\
  \textbf{D95} (2017) 114012},
  \href{http://arxiv.org/abs/1703.07774}{{\normalfont\ttfamily
  arXiv:1703.07774}}\relax
\mciteBstWouldAddEndPuncttrue
\mciteSetBstMidEndSepPunct{\mcitedefaultmidpunct}
{\mcitedefaultendpunct}{\mcitedefaultseppunct}\relax
\EndOfBibitem
\bibitem{Chen:2017sci}
H.-X. Chen {\em et~al.}, \ifthenelse{\boolean{articletitles}}{\emph{{Decay
  properties of $P$-wave charmed baryons from light-cone QCD sum rules}},
  }{}\href{https://doi.org/10.1103/PhysRevD.95.094008}{Phys.\ Rev.\
  \textbf{D95} (2017) 094008},
  \href{http://arxiv.org/abs/1703.07703}{{\normalfont\ttfamily
  arXiv:1703.07703}}\relax
\mciteBstWouldAddEndPuncttrue
\mciteSetBstMidEndSepPunct{\mcitedefaultmidpunct}
{\mcitedefaultendpunct}{\mcitedefaultseppunct}\relax
\EndOfBibitem
\bibitem{Wang:2017hej}
K.-L. Wang, L.-Y. Xiao, X.-H. Zhong, and Q.~Zhao,
  \ifthenelse{\boolean{articletitles}}{\emph{{Understanding the newly observed
  $\Omegares_c$ states through their decays}},
  }{}\href{https://doi.org/10.1103/PhysRevD.95.116010}{Phys.\ Rev.\
  \textbf{D95} (2017) 116010},
  \href{http://arxiv.org/abs/1703.09130}{{\normalfont\ttfamily
  arXiv:1703.09130}}\relax
\mciteBstWouldAddEndPuncttrue
\mciteSetBstMidEndSepPunct{\mcitedefaultmidpunct}
{\mcitedefaultendpunct}{\mcitedefaultseppunct}\relax
\EndOfBibitem
\bibitem{Wang:2017vnc}
W.~Wang and R.-L. Zhu,
  \ifthenelse{\boolean{articletitles}}{\emph{{Interpretation of the newly
  observed $\Omegares_c^0$ resonances}},
  }{}\href{https://doi.org/10.1103/PhysRevD.96.014024}{Phys.\ Rev.\
  \textbf{D96} (2017) 014024},
  \href{http://arxiv.org/abs/1704.00179}{{\normalfont\ttfamily
  arXiv:1704.00179}}\relax
\mciteBstWouldAddEndPuncttrue
\mciteSetBstMidEndSepPunct{\mcitedefaultmidpunct}
{\mcitedefaultendpunct}{\mcitedefaultseppunct}\relax
\EndOfBibitem
\bibitem{Cheng:2017ove}
H.-Y. Cheng and C.-W. Chiang,
  \ifthenelse{\boolean{articletitles}}{\emph{{Quantum numbers of $\Omegares_c$
  states and other charmed baryons}},
  }{}\href{https://doi.org/10.1103/PhysRevD.95.094018}{Phys.\ Rev.\
  \textbf{D95} (2017) 094018},
  \href{http://arxiv.org/abs/1704.00396}{{\normalfont\ttfamily
  arXiv:1704.00396}}\relax
\mciteBstWouldAddEndPuncttrue
\mciteSetBstMidEndSepPunct{\mcitedefaultmidpunct}
{\mcitedefaultendpunct}{\mcitedefaultseppunct}\relax
\EndOfBibitem
\bibitem{Zhao:2017fov}
Z.~Zhao, D.-D. Ye, and A.~Zhang,
  \ifthenelse{\boolean{articletitles}}{\emph{{Hadronic decay properties of
  newly observed $\Omegares_c$ baryons}},
  }{}\href{https://doi.org/10.1103/PhysRevD.95.114024}{Phys.\ Rev.\
  \textbf{D95} (2017) 114024},
  \href{http://arxiv.org/abs/1704.02688}{{\normalfont\ttfamily
  arXiv:1704.02688}}\relax
\mciteBstWouldAddEndPuncttrue
\mciteSetBstMidEndSepPunct{\mcitedefaultmidpunct}
{\mcitedefaultendpunct}{\mcitedefaultseppunct}\relax
\EndOfBibitem
\bibitem{Agaev:2017lip}
S.~S. Agaev, K.~Azizi, and H.~Sundu,
  \ifthenelse{\boolean{articletitles}}{\emph{{Interpretation of the new
  $\Omegares_c^{0}$ states via their mass and width}},
  }{}\href{https://doi.org/10.1140/epjc/s10052-017-4953-z}{Eur.\ Phys.\ J.\
  \textbf{C77} (2017) 395},
  \href{http://arxiv.org/abs/1704.04928}{{\normalfont\ttfamily
  arXiv:1704.04928}}\relax
\mciteBstWouldAddEndPuncttrue
\mciteSetBstMidEndSepPunct{\mcitedefaultmidpunct}
{\mcitedefaultendpunct}{\mcitedefaultseppunct}\relax
\EndOfBibitem
\bibitem{Chen:2017gnu}
B.~Chen and X.~Liu, \ifthenelse{\boolean{articletitles}}{\emph{{New
  $\Omegares_c^0$ baryons discovered by LHCb as the members of $1P$ and $2S$
  states}}, }{}\href{https://doi.org/10.1103/PhysRevD.96.094015}{Phys.\ Rev.\
  \textbf{D96} (2017) 094015},
  \href{http://arxiv.org/abs/1704.02583}{{\normalfont\ttfamily
  arXiv:1704.02583}}\relax
\mciteBstWouldAddEndPuncttrue
\mciteSetBstMidEndSepPunct{\mcitedefaultmidpunct}
{\mcitedefaultendpunct}{\mcitedefaultseppunct}\relax
\EndOfBibitem
\bibitem{Wang:2017zjw}
Z.-G. Wang, \ifthenelse{\boolean{articletitles}}{\emph{{Analysis of $\Omegares
  _c(3000)$, $\Omegares _c(3050)$, $\Omegares _c(3066)$, $\Omegares _c(3090)$
  and $\Omegares _c(3119)$ with QCD sum rules}},
  }{}\href{https://doi.org/10.1140/epjc/s10052-017-4895-5}{Eur.\ Phys.\ J.\
  \textbf{C77} (2017) 325},
  \href{http://arxiv.org/abs/1704.01854}{{\normalfont\ttfamily
  arXiv:1704.01854}}\relax
\mciteBstWouldAddEndPuncttrue
\mciteSetBstMidEndSepPunct{\mcitedefaultmidpunct}
{\mcitedefaultendpunct}{\mcitedefaultseppunct}\relax
\EndOfBibitem
\bibitem{Wang:2017kfr}
K.-L. Wang, Y.-X. Yao, X.-H. Zhong, and Q.~Zhao,
  \ifthenelse{\boolean{articletitles}}{\emph{{Strong and radiative decays of
  the low-lying $S$- and $P$-wave singly heavy baryons}},
  }{}\href{https://doi.org/10.1103/PhysRevD.96.116016}{Phys.\ Rev.\
  \textbf{D96} (2017) 116016},
  \href{http://arxiv.org/abs/1709.04268}{{\normalfont\ttfamily
  arXiv:1709.04268}}\relax
\mciteBstWouldAddEndPuncttrue
\mciteSetBstMidEndSepPunct{\mcitedefaultmidpunct}
{\mcitedefaultendpunct}{\mcitedefaultseppunct}\relax
\EndOfBibitem
\bibitem{Yao:2018jmc}
Y.-X. Yao, K.-L. Wang, and X.-H. Zhong,
  \ifthenelse{\boolean{articletitles}}{\emph{{Strong and radiative decays of
  the low-lying $D$-wave singly heavy baryons}},
  }{}\href{https://doi.org/10.1103/PhysRevD.98.076015}{Phys.\ Rev.\
  \textbf{D98} (2018) 076015},
  \href{http://arxiv.org/abs/1803.00364}{{\normalfont\ttfamily
  arXiv:1803.00364}}\relax
\mciteBstWouldAddEndPuncttrue
\mciteSetBstMidEndSepPunct{\mcitedefaultmidpunct}
{\mcitedefaultendpunct}{\mcitedefaultseppunct}\relax
\EndOfBibitem
\bibitem{Mao:2017wbz}
Q.~Mao {\em et~al.}, \ifthenelse{\boolean{articletitles}}{\emph{{$D$-wave heavy
  baryons of the $SU(3)$ flavor $\mathbf{6}_F$}},
  }{}\href{https://doi.org/10.1103/PhysRevD.96.074021}{Phys.\ Rev.\
  \textbf{D96} (2017) 074021},
  \href{http://arxiv.org/abs/1707.03712}{{\normalfont\ttfamily
  arXiv:1707.03712}}\relax
\mciteBstWouldAddEndPuncttrue
\mciteSetBstMidEndSepPunct{\mcitedefaultmidpunct}
{\mcitedefaultendpunct}{\mcitedefaultseppunct}\relax
\EndOfBibitem
\bibitem{Santopinto:2018ljf}
E.~Santopinto {\em et~al.}, \ifthenelse{\boolean{articletitles}}{\emph{{The
  $\POmega _{ c}$-puzzle solved by means of quark model predictions}},
  }{}\href{https://doi.org/10.1140/epjc/s10052-019-7527-4}{Eur.\ Phys.\ J.\
  \textbf{C79} (2019) 1012},
  \href{http://arxiv.org/abs/1811.01799}{{\normalfont\ttfamily
  arXiv:1811.01799}}\relax
\mciteBstWouldAddEndPuncttrue
\mciteSetBstMidEndSepPunct{\mcitedefaultmidpunct}
{\mcitedefaultendpunct}{\mcitedefaultseppunct}\relax
\EndOfBibitem
\bibitem{Xu:2019kkt}
H.-g. Xu {\em et~al.}, \ifthenelse{\boolean{articletitles}}{\emph{{Study on the
  $\Omegares_{c}^{0}$ states decaying to $\Xi_{c}^{+}K^{-}$ in $pp$ collisions
  at $\sqrt{s}=7$, 13 TeV}},
  }{}\href{https://doi.org/10.1103/PhysRevC.102.054319}{Phys.\ Rev.\
  \textbf{C102} (2020) 054319},
  \href{http://arxiv.org/abs/1912.12905}{{\normalfont\ttfamily
  arXiv:1912.12905}}\relax
\mciteBstWouldAddEndPuncttrue
\mciteSetBstMidEndSepPunct{\mcitedefaultmidpunct}
{\mcitedefaultendpunct}{\mcitedefaultseppunct}\relax
\EndOfBibitem
\bibitem{Hu:2020nkg}
S.~Hu, G.~Meng, and F.~Xu, \ifthenelse{\boolean{articletitles}}{\emph{{Hadronic
  weak decays of the charmed baryon $\Omegares_c$}},
  }{}\href{https://doi.org/10.1103/PhysRevD.101.094033}{Phys.\ Rev.\
  \textbf{D101} (2020) 094033},
  \href{http://arxiv.org/abs/2003.04705}{{\normalfont\ttfamily
  arXiv:2003.04705}}\relax
\mciteBstWouldAddEndPuncttrue
\mciteSetBstMidEndSepPunct{\mcitedefaultmidpunct}
{\mcitedefaultendpunct}{\mcitedefaultseppunct}\relax
\EndOfBibitem
\bibitem{Yang:2021lce}
H.-M. Yang and H.-X. Chen, \ifthenelse{\boolean{articletitles}}{\emph{{P-wave
  charmed baryons of the SU(3) flavor 6F}},
  }{}\href{https://doi.org/10.1103/PhysRevD.104.034037}{Phys.\ Rev.\
  \textbf{D104} (2021) 034037},
  \href{http://arxiv.org/abs/2106.15488}{{\normalfont\ttfamily
  arXiv:2106.15488}}\relax
\mciteBstWouldAddEndPuncttrue
\mciteSetBstMidEndSepPunct{\mcitedefaultmidpunct}
{\mcitedefaultendpunct}{\mcitedefaultseppunct}\relax
\EndOfBibitem
\bibitem{RamirezMorales:2022lsy}
A.~Ramirez~Morales, \ifthenelse{\boolean{articletitles}}{\emph{{Strong decay
  widths and mass spectra of charmed baryons}},
  }{}\href{https://doi.org/10.22323/1.414.0913}{PoS \textbf{ICHEP2022} (2022)
  913}\relax
\mciteBstWouldAddEndPuncttrue
\mciteSetBstMidEndSepPunct{\mcitedefaultmidpunct}
{\mcitedefaultendpunct}{\mcitedefaultseppunct}\relax
\EndOfBibitem
\bibitem{Padmanath:2017lng}
M.~Padmanath and N.~Mathur, \ifthenelse{\boolean{articletitles}}{\emph{{Quantum
  numbers of recently discovered $\Omegares^{0}_{c}$ baryons from lattice
  QCD}}, }{}\href{https://doi.org/10.1103/PhysRevLett.119.042001}{Phys.\ Rev.\
  Lett.\  \textbf{119} (2017) 042001},
  \href{http://arxiv.org/abs/1704.00259}{{\normalfont\ttfamily
  arXiv:1704.00259}}\relax
\mciteBstWouldAddEndPuncttrue
\mciteSetBstMidEndSepPunct{\mcitedefaultmidpunct}
{\mcitedefaultendpunct}{\mcitedefaultseppunct}\relax
\EndOfBibitem
\bibitem{Debastiani:2017ewu}
V.~R. Debastiani, J.~M. Dias, W.~H. Liang, and E.~Oset,
  \ifthenelse{\boolean{articletitles}}{\emph{{Molecular $\Omegares_c$ states
  generated from coupled meson-baryon channels}},
  }{}\href{https://doi.org/10.1103/PhysRevD.97.094035}{Phys.\ Rev.\
  \textbf{D97} (2018) 094035},
  \href{http://arxiv.org/abs/1710.04231}{{\normalfont\ttfamily
  arXiv:1710.04231}}\relax
\mciteBstWouldAddEndPuncttrue
\mciteSetBstMidEndSepPunct{\mcitedefaultmidpunct}
{\mcitedefaultendpunct}{\mcitedefaultseppunct}\relax
\EndOfBibitem
\bibitem{Montana:2017kjw}
G.~Monta\~na, A.~Feijoo, and A.~Ramos,
  \ifthenelse{\boolean{articletitles}}{\emph{{A meson-baryon molecular
  interpretation for some $\Omegares_{c}$ excited states}},
  }{}\href{https://doi.org/10.1140/epja/i2018-12498-1}{Eur.\ Phys.\ J.\
  \textbf{A54} (2018) 64},
  \href{http://arxiv.org/abs/1709.08737}{{\normalfont\ttfamily
  arXiv:1709.08737}}\relax
\mciteBstWouldAddEndPuncttrue
\mciteSetBstMidEndSepPunct{\mcitedefaultmidpunct}
{\mcitedefaultendpunct}{\mcitedefaultseppunct}\relax
\EndOfBibitem
\bibitem{Ramos:2020bgs}
A.~Ramos, A.~Feijoo, Q.~Llorens, and G.~Monta\~na,
  \ifthenelse{\boolean{articletitles}}{\emph{{The molecular nature of some
  exotic hadrons}}, }{}\href{https://doi.org/10.1007/s00601-020-01566-0}{Few
  Body Syst.\  \textbf{61} (2020) 34},
  \href{http://arxiv.org/abs/2009.04367}{{\normalfont\ttfamily
  arXiv:2009.04367}}\relax
\mciteBstWouldAddEndPuncttrue
\mciteSetBstMidEndSepPunct{\mcitedefaultmidpunct}
{\mcitedefaultendpunct}{\mcitedefaultseppunct}\relax
\EndOfBibitem
\bibitem{An:2017lwg}
C.~S. An and H.~Chen, \ifthenelse{\boolean{articletitles}}{\emph{{Observed
  $\Omegares_{c}^{0}$ resonances as pentaquark states}},
  }{}\href{https://doi.org/10.1103/PhysRevD.96.034012}{Phys.\ Rev.\
  \textbf{D96} (2017) 034012},
  \href{http://arxiv.org/abs/1705.08571}{{\normalfont\ttfamily
  arXiv:1705.08571}}\relax
\mciteBstWouldAddEndPuncttrue
\mciteSetBstMidEndSepPunct{\mcitedefaultmidpunct}
{\mcitedefaultendpunct}{\mcitedefaultseppunct}\relax
\EndOfBibitem
\bibitem{Wang_2021}
H.-J. Wang, Z.-Y. Di, and Z.-G. Wang,
  \ifthenelse{\boolean{articletitles}}{\emph{{Analysis of the excited
  $\Omegares_c$ states as the $\frac{1}{2}^{\pm}$ pentaquark states with QCD
  sum rules}},
  }{}\href{https://doi.org/10.1088/1572-9494/abc7b1}{Communications in
  Theoretical Physics \textbf{73} (2021) 035201}\relax
\mciteBstWouldAddEndPuncttrue
\mciteSetBstMidEndSepPunct{\mcitedefaultmidpunct}
{\mcitedefaultendpunct}{\mcitedefaultseppunct}\relax
\EndOfBibitem
\bibitem{LHCb-DP-2012-004}
R.~Aaij {\em et~al.}, \ifthenelse{\boolean{articletitles}}{\emph{{The \lhcb
  trigger and its performance in 2011}},
  }{}\href{https://doi.org/10.1088/1748-0221/8/04/P04022}{JINST \textbf{8}
  (2013) P04022}, \href{http://arxiv.org/abs/1211.3055}{{\normalfont\ttfamily
  arXiv:1211.3055}}\relax
\mciteBstWouldAddEndPuncttrue
\mciteSetBstMidEndSepPunct{\mcitedefaultmidpunct}
{\mcitedefaultendpunct}{\mcitedefaultseppunct}\relax
\EndOfBibitem
\bibitem{LHCb-DP-2019-001}
R.~Aaij {\em et~al.}, \ifthenelse{\boolean{articletitles}}{\emph{{Design and
  performance of the LHCb trigger and full real-time reconstruction in Run 2 of
  the LHC}}, }{}\href{https://doi.org/10.1088/1748-0221/14/04/P04013}{JINST
  \textbf{14} (2019) P04013},
  \href{http://arxiv.org/abs/1812.10790}{{\normalfont\ttfamily
  arXiv:1812.10790}}\relax
\mciteBstWouldAddEndPuncttrue
\mciteSetBstMidEndSepPunct{\mcitedefaultmidpunct}
{\mcitedefaultendpunct}{\mcitedefaultseppunct}\relax
\EndOfBibitem
\bibitem{Alves:2008zz}
LHCb collaboration, A.~A. Alves~Jr.\ {\em et~al.},
  \ifthenelse{\boolean{articletitles}}{\emph{{The \lhcb detector at the LHC}},
  }{}\href{https://doi.org/10.1088/1748-0221/3/08/S08005}{JINST \textbf{3}
  (2008) S08005}\relax
\mciteBstWouldAddEndPuncttrue
\mciteSetBstMidEndSepPunct{\mcitedefaultmidpunct}
{\mcitedefaultendpunct}{\mcitedefaultseppunct}\relax
\EndOfBibitem
\bibitem{LHCb-DP-2014-002}
LHCb collaboration, R.~Aaij {\em et~al.},
  \ifthenelse{\boolean{articletitles}}{\emph{{LHCb detector performance}},
  }{}\href{https://doi.org/10.1142/S0217751X15300227}{Int.\ J.\ Mod.\ Phys.\
  \textbf{A30} (2015) 1530022},
  \href{http://arxiv.org/abs/1412.6352}{{\normalfont\ttfamily
  arXiv:1412.6352}}\relax
\mciteBstWouldAddEndPuncttrue
\mciteSetBstMidEndSepPunct{\mcitedefaultmidpunct}
{\mcitedefaultendpunct}{\mcitedefaultseppunct}\relax
\EndOfBibitem
\bibitem{Sjostrand:2007gs}
T.~Sj\"{o}strand, S.~Mrenna, and P.~Skands,
  \ifthenelse{\boolean{articletitles}}{\emph{{A brief introduction to PYTHIA
  8.1}}, }{}\href{https://doi.org/10.1016/j.cpc.2008.01.036}{Comput.\ Phys.\
  Commun.\  \textbf{178} (2008) 852},
  \href{http://arxiv.org/abs/0710.3820}{{\normalfont\ttfamily
  arXiv:0710.3820}}\relax
\mciteBstWouldAddEndPuncttrue
\mciteSetBstMidEndSepPunct{\mcitedefaultmidpunct}
{\mcitedefaultendpunct}{\mcitedefaultseppunct}\relax
\EndOfBibitem
\bibitem{LHCb:2011dpk}
LHCb collaboration, I.~Belyaev {\em et~al.},
  \ifthenelse{\boolean{articletitles}}{\emph{{Handling of the generation of
  primary events in Gauss, the LHCb simulation framework}},
  }{}\href{https://doi.org/10.1088/1742-6596/331/3/032047}{J.\ Phys.\ Conf.\
  Ser.\  \textbf{331} (2011) 032047}\relax
\mciteBstWouldAddEndPuncttrue
\mciteSetBstMidEndSepPunct{\mcitedefaultmidpunct}
{\mcitedefaultendpunct}{\mcitedefaultseppunct}\relax
\EndOfBibitem
\bibitem{Lange:2001uf}
D.~J. Lange, \ifthenelse{\boolean{articletitles}}{\emph{{The EvtGen particle
  decay simulation package}},
  }{}\href{https://doi.org/10.1016/S0168-9002(01)00089-4}{Nucl.\ Instrum.\
  Meth.\  \textbf{A462} (2001) 152}\relax
\mciteBstWouldAddEndPuncttrue
\mciteSetBstMidEndSepPunct{\mcitedefaultmidpunct}
{\mcitedefaultendpunct}{\mcitedefaultseppunct}\relax
\EndOfBibitem
\bibitem{Allison:2006ve}
Geant4 collaboration, J.~Allison {\em et~al.},
  \ifthenelse{\boolean{articletitles}}{\emph{{Geant4 developments and
  applications}}, }{}\href{https://doi.org/10.1109/TNS.2006.869826}{IEEE
  Trans.\ Nucl.\ Sci.\  \textbf{53} (2006) 270}\relax
\mciteBstWouldAddEndPuncttrue
\mciteSetBstMidEndSepPunct{\mcitedefaultmidpunct}
{\mcitedefaultendpunct}{\mcitedefaultseppunct}\relax
\EndOfBibitem
\bibitem{Clemencic:2011zza}
LHCb collaboration, M.~Clemencic {\em et~al.},
  \ifthenelse{\boolean{articletitles}}{\emph{{The LHCb simulation application,
  Gauss: Design, evolution and experience}},
  }{}\href{https://doi.org/10.1088/1742-6596/331/3/032023}{J.\ Phys.\ Conf.\
  Ser.\  \textbf{331} (2011) 032023}\relax
\mciteBstWouldAddEndPuncttrue
\mciteSetBstMidEndSepPunct{\mcitedefaultmidpunct}
{\mcitedefaultendpunct}{\mcitedefaultseppunct}\relax
\EndOfBibitem
\bibitem{PDG2022}
Particle Data Group, R.~L. Workman and Others,
  \ifthenelse{\boolean{articletitles}}{\emph{{Review of particle physics}},
  }{}\href{https://doi.org/10.1093/ptep/ptac097}{PTEP \textbf{2022} (2022)
  083C01}\relax
\mciteBstWouldAddEndPuncttrue
\mciteSetBstMidEndSepPunct{\mcitedefaultmidpunct}
{\mcitedefaultendpunct}{\mcitedefaultseppunct}\relax
\EndOfBibitem
\bibitem{BBDT}
V.~V. Gligorov and M.~Williams,
  \ifthenelse{\boolean{articletitles}}{\emph{{Efficient, reliable and fast
  high-level triggering using a bonsai boosted decision tree}},
  }{}\href{https://doi.org/10.1088/1748-0221/8/02/P02013}{JINST \textbf{8}
  (2013) P02013}, \href{http://arxiv.org/abs/1210.6861}{{\normalfont\ttfamily
  arXiv:1210.6861}}\relax
\mciteBstWouldAddEndPuncttrue
\mciteSetBstMidEndSepPunct{\mcitedefaultmidpunct}
{\mcitedefaultendpunct}{\mcitedefaultseppunct}\relax
\EndOfBibitem
\bibitem{TMVA4}
A.~Hoecker {\em et~al.}, \ifthenelse{\boolean{articletitles}}{\emph{{TMVA 4 ---
  toolkit for multivariate data analysis with ROOT, users guide}},
  }{}\href{http://arxiv.org/abs/physics/0703039}{{\normalfont\ttfamily
  arXiv:physics/0703039}}\relax
\mciteBstWouldAddEndPuncttrue
\mciteSetBstMidEndSepPunct{\mcitedefaultmidpunct}
{\mcitedefaultendpunct}{\mcitedefaultseppunct}\relax
\EndOfBibitem
\bibitem{Cowan:2016tnm}
G.~A. Cowan, D.~C. Craik, and M.~D. Needham,
  \ifthenelse{\boolean{articletitles}}{\emph{{RapidSim: an application for the
  fast simulation of heavy-quark hadron decays}},
  }{}\href{https://doi.org/10.1016/j.cpc.2017.01.029}{Comput.\ Phys.\ Commun.\
  \textbf{214} (2017) 239},
  \href{http://arxiv.org/abs/1612.07489}{{\normalfont\ttfamily
  arXiv:1612.07489}}\relax
\mciteBstWouldAddEndPuncttrue
\mciteSetBstMidEndSepPunct{\mcitedefaultmidpunct}
{\mcitedefaultendpunct}{\mcitedefaultseppunct}\relax
\EndOfBibitem
\bibitem{Blatt:1952ije}
J.~M. Blatt and V.~F. Weisskopf, {\em {Theoretical nuclear physics}},
  \href{https://doi.org/10.1007/978-1-4612-9959-2}{ Springer, New York,
  1952}\relax
\mciteBstWouldAddEndPuncttrue
\mciteSetBstMidEndSepPunct{\mcitedefaultmidpunct}
{\mcitedefaultendpunct}{\mcitedefaultseppunct}\relax
\EndOfBibitem
\bibitem{LHCb-PAPER-2012-048}
LHCb collaboration, R.~Aaij {\em et~al.},
  \ifthenelse{\boolean{articletitles}}{\emph{{Measurements of the \Lb, \Xibm,
  and \Omegab baryon masses}},
  }{}\href{https://doi.org/10.1103/PhysRevLett.110.182001}{Phys.\ Rev.\ Lett.\
  \textbf{110} (2013) 182001},
  \href{http://arxiv.org/abs/1302.1072}{{\normalfont\ttfamily
  arXiv:1302.1072}}\relax
\mciteBstWouldAddEndPuncttrue
\mciteSetBstMidEndSepPunct{\mcitedefaultmidpunct}
{\mcitedefaultendpunct}{\mcitedefaultseppunct}\relax
\EndOfBibitem
\bibitem{LHCb-PAPER-2013-011}
LHCb collaboration, R.~Aaij {\em et~al.},
  \ifthenelse{\boolean{articletitles}}{\emph{{Precision measurement of \D meson
  mass differences}}, }{}\href{https://doi.org/10.1007/JHEP06(2013)065}{JHEP
  \textbf{06} (2013) 065},
  \href{http://arxiv.org/abs/1304.6865}{{\normalfont\ttfamily
  arXiv:1304.6865}}\relax
\mciteBstWouldAddEndPuncttrue
\mciteSetBstMidEndSepPunct{\mcitedefaultmidpunct}
{\mcitedefaultendpunct}{\mcitedefaultseppunct}\relax
\EndOfBibitem
\bibitem{LHCB-PAPER-2014-048}
LHCb collaboration, R.~Aaij {\em et~al.},
  \ifthenelse{\boolean{articletitles}}{\emph{{Precision measurement of the mass
  and lifetime of the \Xibm baryon}},
  }{}\href{https://doi.org/10.1103/PhysRevLett.113.242002}{Phys.\ Rev.\ Lett.\
  \textbf{113} (2014) 242002},
  \href{http://arxiv.org/abs/1409.8568}{{\normalfont\ttfamily
  arXiv:1409.8568}}\relax
\mciteBstWouldAddEndPuncttrue
\mciteSetBstMidEndSepPunct{\mcitedefaultmidpunct}
{\mcitedefaultendpunct}{\mcitedefaultseppunct}\relax
\EndOfBibitem
\end{mcitethebibliography}
\end{document}